\documentclass[11pt,twoside]{article}

\usepackage{asp2006}
\usepackage{epsf}
\usepackage{psfig}
\usepackage{lscape}

\begin{document}
\title{A Photometric Transit Search for Planets around Cool Stars from the Italian Alps: Results from a Feasibility Study}   
\author{Mario Damasso, Paolo Calcidese, Andrea Bernagozzi and Enzo Bertolini}   
\affil{Astronomical Observatory of the Autonomous Region of the Aosta Valley, Loc. Lignan 39, Nus (Aosta) - Italy}    
\author{Paolo Giacobbe}
\affil{Dept. of Physics, University of Turin, Via Giuria 1, I-10125, Turin - Italy}
\author{Mario G. Lattanzi, Richard Smart and Alessandro Sozzetti}
\affil{INAF-Astronomical Observatory of Turin, Strada Osservatorio, 20
I-10025 Pino Torinese (Turin) - Italy}

\begin{abstract}
A feasibility study was carried out at the Astronomical Observatory of the Autonomous Region of the Aosta Valley demonstrating that it is a well-poised site to conduct an upcoming observing campaign aimed at detecting small-size (R$\leq$R{\tiny Neptune\/}) transiting planets around nearby cool M dwarf stars. Three known transiting planet systems were monitored from May to August 2009 with a 25 cm f/3.8 Maksutov telescope. We reached seeing-independent, best-case photometric RMS less than 0.003 mag for stars with V$\leq$13, with a median RMS of 0.006 mag for the whole observing period.
\end{abstract}

\section{Introduction}
The Astronomical Observatory of the Autonomous Region of the Aosta Valley (OAVdA; 45.7895\deg  N, 7.47833\deg E) has been identified as a potential site for hosting a photometric transit search for low-mass, small-size planets around nearby cool M dwarf stars. We carried out a study to gauge the near-term and long-term photometric precision achievable, observing known transiting systems in a range of transit depths under variable seeing and sky transparency conditions.

\section{Instrumentation and Methodology}   
We used a 25 cm Maksutov f/3.8 reflector telescope equipped with a CCD Moravian G2-3200ME and an R filter centered at 610 nm (field of view: 52.10 x 35.11 arcmin$^{2}$; plate scale: 1.43 arcsec/pix; QE$\approx$87\% at 610 nm). The seeing was monitored each night using a Hartmann mask and the DIMM technique. It varied in the range 1-3 arcsec (median $\approx$1.7 arcsec). Data reduction and ensemble differential aperture photometry ($\approx$100 reference objects) were carried out with the version 1.0 of an automated, IDL-based pipeline we have developed.

\begin{figure}
\plottwo{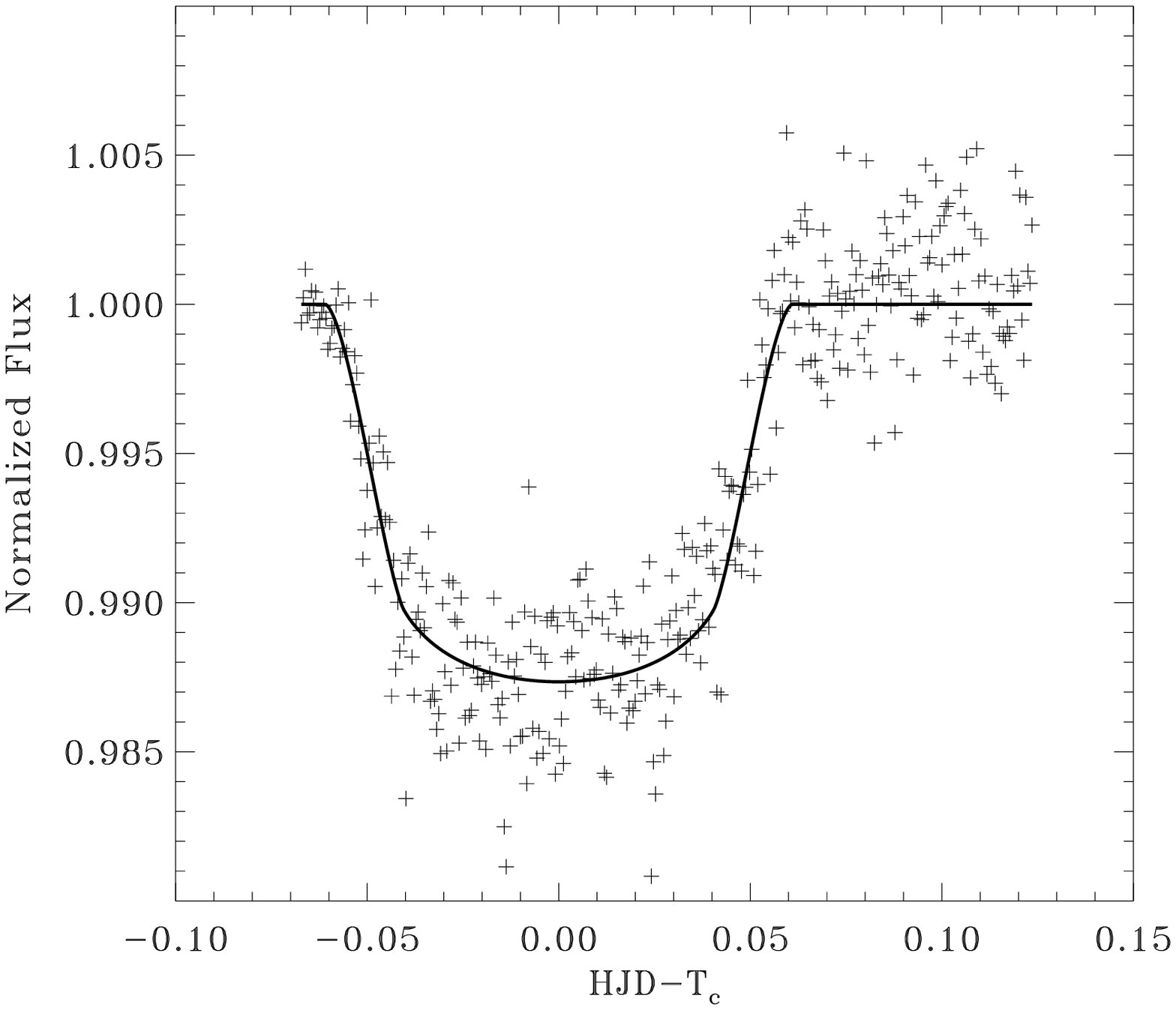}{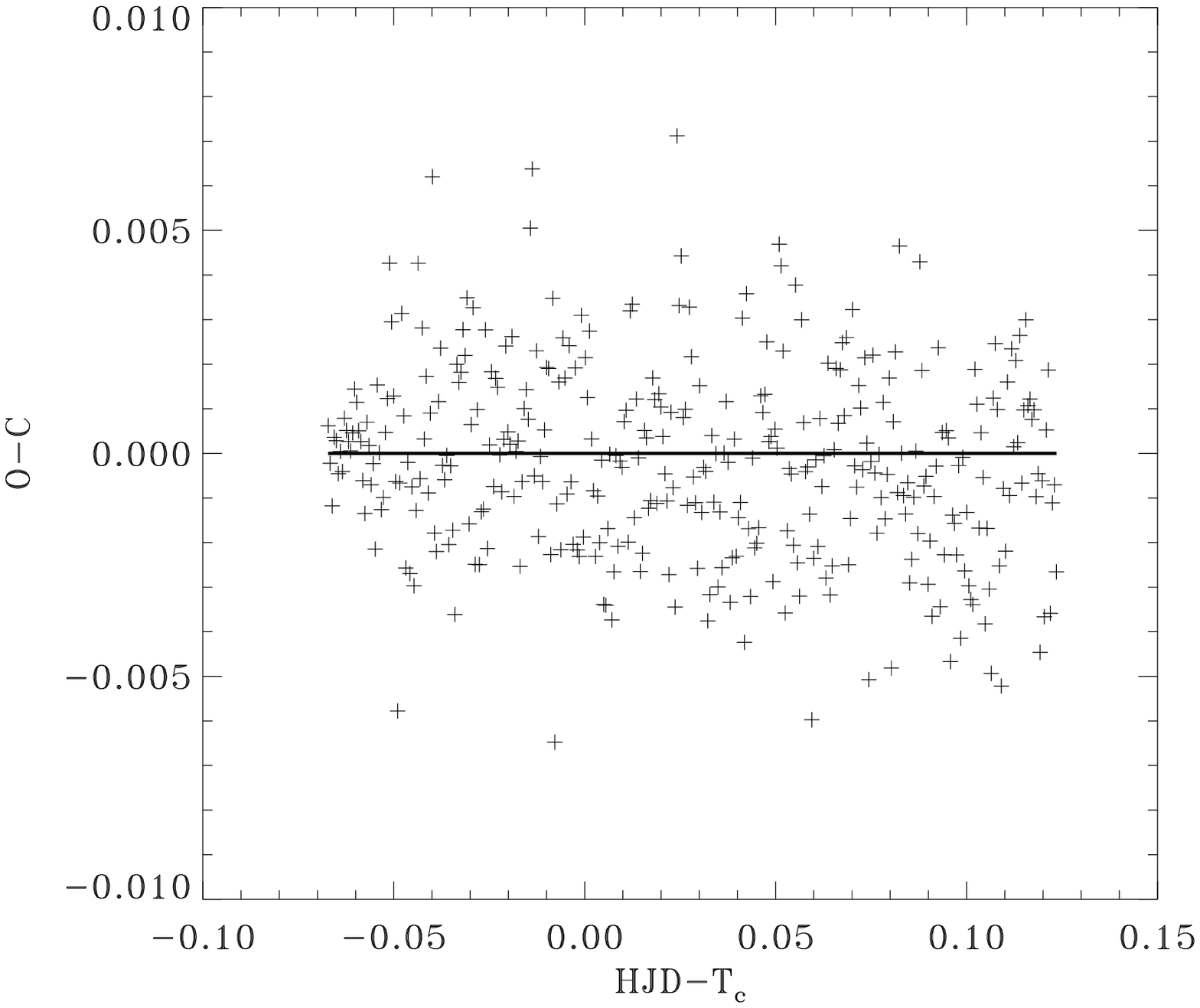}
\vspace{-0.005cm}
\caption{{\itshape Left:\/} WASP-3b transit light curve (28-29 July 2009; expos. 35 s; T{\tiny C\/}=55041.411) and best-fit curve. {\itshape Right:\/} residuals of the fitted curve.}
\end{figure}

\section{Selected Results}
In May-August 2009 we monitored three transiting systems: WASP-3 \citep{Dam-Pol07}, HAT-P-7 \citep{Dam-Pal08} and Gliese 436 \citep{Dam-But04}. WASP-3, a 1.24 M{\tiny Sun\/} star (V=10.64) hosting a 1.76 M{\tiny Jupiter\/} planet (period=1.846834 days), was monitored for 19 nights, often reaching a photometric RMS below 0.003 mag for stars with V$\leq$13. Fig.1 shows one of the observed WASP-3b transits and the best-fit model is based on the formalism of \Citet{Dam-Mand02}, assuming quadratic limb darkening. We got: b=0.665$\pm$0.003 (impact parameter), i=81.24$\pm$0.06 degrees (orbital inclination), r=0.1091$\pm$0.0006 (planet-to-star radius ratio), T{\tiny C\/}=55041.411$\pm$0.029 HJD (time of mid transit). The median RMS for the whole observing period is 0.006 mag (V$\leq$13).

\section{Conclusions and Future Work}
The feasibility study demonstrated that the OAVdA is a promising site for a long-term photometric survey aimed to detect transiting low-mass, small-size planets. We are planning to set up a system of five 40 cm identical telescopes for an high-precision observing campaign focused on several hundreds nearby M dwarfs, improving the longitudinal coverage of similar programs already ongoing, such as the Arizona-based MEarth project \Citep{Dam-Nut08}.

\acknowledgements M.D., P.C. and A.B. benefit of a grant provided by the European Union, the Autonomous Region of the Aosta Valley and the Italian Department for Work, Health and Pensions.


\begin{thebibliography}{}
\bibitem[Butler et al.(2004)]{Dam-But04}Butler, P. et al. 2004, ApJ, 617, L580
\bibitem[Mandel \& Agol(2002)]{Dam-Mand02}Mandel, K. \& Agol, E. 2002, ApJ, 580, L171
\bibitem[Nutzman \& Charbonneau(2008)]{Dam-Nut08}Nutzman, P. \& Charbonneau, D. 2008, PASP, 120, 317
\bibitem[P\'al et al.(2008)]{Dam-Pal08}P\'al, A. et al. 2008 ApJ, 680, 2, 1450 
\bibitem[Pollacco et al.(2007)]{Dam-Pol07}Pollacco, D. et al. 2007, MNRAS, 385, 1576
\end{thebibliography}
\end{document}